\documentclass[11pt]{article}
\usepackage{amsfonts}
\usepackage{amssymb}
\usepackage{amsmath}
\usepackage{graphicx}

\newcommand{\ket}[1]{|#1\rangle}

\newcommand{\op}[1]{\operatorname{#1}}

\newtheorem{theorem}{Theorem}

\newtheorem{definition}{Definition}[section]

\begin{document}

\title{\Large\bf Quantum algorithms for solvable groups}

\author{
John Watrous\thanks{Partially supported by Canada's NSERC.}\\
Department of Computer Science\\
University of Calgary\\
Calgary, Alberta, Canada\\
jwatrous@cpsc.ucalgary.ca
}

\date{January 26, 2001}

\maketitle

\thispagestyle{empty}

\begin{abstract}
In this paper we give a polynomial-time quantum algorithm for computing orders
of solvable groups.
Several other problems, such as testing membership in solvable groups,
testing equality of subgroups in a given solvable group, and testing normality
of a subgroup in a given solvable group, reduce to computing orders of
solvable groups and therefore admit polynomial-time quantum algorithms as well.
Our algorithm works in the setting of black-box groups, wherein
none of these problems can be computed classically in polynomial time.
As an important byproduct, our algorithm is able to produce a pure quantum
state that is uniform over the elements in any chosen subgroup of a solvable
group, which yields a natural way to apply existing quantum algorithms
to factor groups of solvable groups.
\end{abstract}


\section{Introduction}
\label{sec:introduction}

The focus of this paper is on quantum algorithms for group-theoretic problems.
Specifically we consider finite solvable groups, and give a polynomial-time
quantum algorithm for computing orders of solvable groups.
Naturally this algorithm yields polynomial-time quantum algorithms for
testing membership in solvable groups and several other related problems that
reduce to computing orders of solvable groups.
Our algorithm is also able to produce a uniform pure state over the elements
in any chosen subgroup of a solvable groups, which yields a natural way
of applying certain quantum algorithms to factor groups of solvable groups.
For instance, we describe a method by which existing quantum algorithms for
abelian groups may be applied to abelian factor groups of solvable groups,
despite the fact that the factor groups generally do not satisfy an important
requirement of the existing quantum algorithms---namely, that elements have
unique, succinct classical representations.

We will be working within the context of {\em black-box groups}, wherein
elements are uniquely encoded by strings of some given length $n$ and the
group operations are performed by a black-box (or {\em group oracle}) at
unit cost.
Black-box groups were introduced by Babai and Szemer\'edi \cite{BabaiS84}
in 1984 and have since been studied extensively
\cite{ArvindV97,Babai91,Babai92,Babai97,BabaiB99,BabaiC+95}.
Any efficient algorithm that works in the context of black-box groups of
course remains efficient whenever the group oracle can be replaced by an
efficient procedure for computing the group operations.
In the black-box group setting it is provably impossible to compute order
classically in polynomial time, even in the more restricted case that groups
in question are abelian \cite{BabaiS84}.

Essentially all previously identified problems for which quantum algorithms
offer exponential speed-up over the best known classical algorithms can be
stated as problems regarding abelian groups.
In 1994, Shor \cite{Shor97} presented polynomial time quantum algorithms for
integer factoring and computing discrete logarithms, and these algorithms
generalize in a natural way to the setting of finite groups.
Specifically, given elements $g$ and $h$ in some finite group $G$ it is
possible, in quantum polynomial time, to find the smallest positive integer
$k$ such that $h = g^k = g\cdot g \cdots g$ ($k$ times), provided there exists
such a $k$.
In case $h$ is the identity one obtains the order of $g$, to which there is a
randomized polynomial-time reduction from factoring when the group is the
multiplicative group of integers modulo the integer $n$ to be factored.
It should be noted that while the group $G$ need not necessarily be abelian
for these algorithms to work, we may view the algorithms as taking place in
the abelian group generated by $g$.

Shor's algorithms for integer factoring and discrete logarithms were
subsequently cast in a different group-theoretic framework by
Kitaev \cite{Kitaev95,Kitaev97}.
This framework involves a problem called the Abelian Stabilizer Problem,
which may be informally stated as follows.
Let $k$ and $n$ be positive integers, and consider some group action of the
additive abelian group $\mathbb{Z}^k$ on a set $X\subseteq\Sigma^n$, where
the group action can be computed efficiently.
The problem, which can be solved in quantum polynomial time, is to compute a
basis (in $\mathbb{Z}^k$) of the stabilizer $(\mathbb{Z}^k)_x$ of a given
$x\in X$.
Appropriate choice of the group action allows one to solve order finding
and discrete logarithms for any finite group as above.
In this case, the group $G$ in question corresponds to the set $X$
(meaning that elements of $X$ are unique representations of elements of
$G$), and the group action of $\mathbb{Z}^k$ on $X$ depends on the group
structure of~$G$.

Kitaev's approach was further generalized by Brassard and H{\o}yer
\cite{BrassardH97}, who formulated the Hidden Subgroup Problem.
(See also H{\o}yer \cite{Hoyer00} and Mosca and Ekert \cite{MoscaE99}.)
The Hidden Subgroup Problem may be informally stated as follows.
Given a finitely generated group $G$ and an efficiently computable function
$f$ from $G$ to some finite set $X$ such that $f$ is constant and distinct on
left-cosets of a subgroup $H$ of finite index, find a generating set for $H$.
Mosca and Ekert showed that Deutsch's Problem~\cite{Deutsch85}, Simon's
Problem~\cite{Simon97}, order finding and computing discrete
logarithms~\cite{Shor97}, finding hidden linear functions~\cite{BonehL95},
testing self-shift-equivalence of polynomials~\cite{Grigoriev96}, and the
Abelian Stabilizer Problem \cite{Kitaev95,Kitaev97} can all be solved in
polynomial time within the framework of the Hidden Subgroup Problem.
In the black-box group setting, the Hidden Subgroup Problem can be solved in
quantum polynomial time whenever $G$ is abelian, as demonstrated by
Mosca \cite{Mosca99}.
Mosca also proved that several other interesting group-theoretic problems
regarding abelian black-box groups can be reduced to the Hidden Subgroup
Problem, and thus can be computed in quantum polynomial time as well.
For instance, given a collection of generators for a finite abelian
black-box group, one can find the order of the group, and in fact one can
decompose the group into a direct product of cyclic subgroups of prime power
order, in polynomial time.\footnote{
This is particularly interesting from the standpoint of algebraic number
theory since, assuming the Generalized Riemann Hypothesis, it follows that
there is a polynomial-time quantum algorithm for computing class numbers of
quadratic number fields.
As there exists a reduction from factoring to the problem of computing class
numbers for quadratic number fields---again assuming the Generalized Riemann
Hypothesis---while no reduction in the other direction is known, the problem
of computing class numbers is often considered as a candidate for a problem
harder than integer factoring.
See Cohen~\cite{Cohen93} for further information about computing in class
groups.}
(See also Cheung and Mosca \cite{CheungM00} for further details.)

The Hidden Subgroup Problem has been considered in the non-abelian case,
although with limited success (see, for instance, Ettinger and
H{\o}yer~\cite{EttingerH99}, Ettinger, H{\o}yer, and Knill~\cite{EttingerH+99},
R\"otteler and Beth~\cite{RoettelerB99}, and Hallgren, Russell, and
Ta-Shma~\cite{HallgrenR+00}).
No polynomial-time algorithm for the Hidden Subgroup Problem is known for
any class of non-abelian groups except for a special class of groups based on
wreath products considered by R\"otteler and Beth.
The Non-abelian Hidden Subgroup Problem is of particular interest as it
relates to the Graph Isomorphism Problem; Graph Isomorphism reduces to a
special case of the Hidden Subgroup Problem in which the groups in question
are the symmetric groups.
Beals \cite{Beals97} has shown that quantum analogues of Fourier transforms
over symmetric groups can be performed in polynomial time, although thus far
this has not proven to be helpful for solving the Graph Isomorphism Problem.

In this paper we move away from the Hidden Subgroup Problem and consider
other group-theoretic problems for non-abelian groups---in particular we
consider solvable groups.
Our main algorithm finds the order of a given solvable group and, as an
important byproduct, produces a quantum state that approximates a uniform
superposition over the elements of the given group.

\begin{theorem}
\label{thm:main}
There exists a quantum algorithm operating as follows (relative to an
arbitrary group oracle).
Given generators $g_1,\ldots,g_k$ such that $G=\langle g_1,\ldots,g_k\rangle$
is solvable, the algorithm outputs the order of $G$ with probability of error
bounded by $\varepsilon$ in time polynomial in \mbox{$n+\log (1/\varepsilon)$}
(where $n$ is the length of the strings representing the generators).
Moreover, the algorithm produces a quantum state $\rho$ that approximates
the pure state $\ket{G} = |G|^{-1/2}\sum_{g\in G}\ket{g}$ with accuracy
$\varepsilon$ (in the trace norm metric).
\end{theorem}

Several other problems reduce to the problem of computing orders of solvable
groups, including membership testing in solvable groups, testing equality of
subgroups in a given solvable group, and testing that a given subgroup of some
solvable group is normal.
Thus, these problems can be solved in quantum polynomial time as well.

Since any subgroup of a solvable group is solvable, our algorithm can be
applied to any subgroup $H$ of a solvable group $G$ in order to obtain a close
approximation to the state $\ket{H}$.
The main application of being able to efficiently prepare uniform
superpositions over subgroups of solvable groups is that it gives us a simple
way to apply existing quantum algorithms for abelian groups to abelian factor
groups of solvable groups, despite the fact that we do not have unique
classical representations for elements in these factor groups.
This method discussed further in Section~\ref{sec:other}.

Arvind and Vinodchandran \cite{ArvindV97} have shown that several problems
regarding solvable groups, including membership testing and order verification,
are low for the complexity class PP, which means that an oracle for these
problems is useless for PP computations.
Fortnow and Rogers \cite{FortnowR99} proved that any problem in BQP is low
for PP, and thus we have obtained an alternate proof that membership testing
and order verification for solvable groups are both low for PP.
It is left open whether some of the other problems proved low for PP by Arvind
and Vinodchandran have polynomial-time quantum algorithms.
An interesting example of such a problem is testing whether two solvable
groups have a nontrivial intersection.

The remainder of this paper has the following organization.
In Section~\ref{sec:preliminaries} we review necessary background information
for this paper, including a discussion of black-box groups in the context of
quantum circuits and other information regarding computational group theory.
Section~\ref{sec:algorithm} describes our quantum algorithm for finding
the order of a solvable group as stated in Theorem~\ref{thm:main}, and
Section~\ref{sec:other} discusses other problems that can be solved by
adapting this algorithm.
We conclude with Section~\ref{sec:conclusion}, which mentions some open
problems relating to this paper.

\subsubsection*{Correction to earlier version}

In an earlier version of this paper it was claimed that our algorithm
could be used to test isomorphism of two solvable groups.
However, this claim was based on an incorrect assumption regarding
solvable groups (specifically that if the corresponding factor groups in the
derived series of two solvable groups are isomorphic, then the groups
themselves are necessarily isomorphic).
Thus, we currently do not have a polynomial-time quantum algorithm for testing
isomorphism of solvable groups.
We thank Miklos Santha for bring this error to our attention.


\section{Preliminaries}
\label{sec:preliminaries}

In this section we review information regarding computational group theory
that is required for the remainder of the paper.
We assume the reader is familiar with the theory of quantum computation,
and specifically with the quantum circuit model, so we will not review this
model further except to discuss black-box groups in the context of quantum
circuits.
The reader not familiar with quantum circuits is referred to
Nielsen and Chuang \cite{NielsenC00}.
We also assume the reader is familiar with the basic concepts of group
theory (see, for example, Isaacs~\cite{Isaacs94}).

Given a group $G$ and elements $g,h\in G$ we define the {\em commutator} of
$g$ and $h$, denoted $[g,h]$, as $[g,h] = g^{-1}h^{-1}gh$, and for any two
subgroups $H,K\leq G$ we write $[H,K]$ to denote the subgroup of $G$ generated
by all commutators $[h,k]$ with $h\in H$ and $k\in K$.
The {\em derived subgroup} of $G$ is $G' = [G,G]$, and in general we write
$G^{(0)}=G, G^{(1)}=G',G^{(2)}=(G')',\ldots,G^{(j)}=(G^{(j-1)})'$, etc.
A group $G$ is said to be {\em solvable} if $G^{(m)} = \{1\}$ (the group
consisting of just one element) for some value of $m$.
Every abelian group is solvable, since $G^{(1)} = \{1\}$ in this case, but it
is not necessarily the case that a given solvable group is abelian (for
example, $S_3$, the symmetric group on 3 symbols, is solvable but not
abelian).
On the other hand many groups are not solvable (for example, $S_n$ is not
solvable whenever $n\geq 5$).
An equivalent way to define what it means for a (finite) group to be solvable
is as follows.
A finite group $G$ is solvable if there exist elements $g_1,\ldots,g_m\in G$
such that if we define $H_j = \langle g_1,\ldots,g_j\rangle$ for each $j$, then
$
\{1\} = H_0\,\triangleleft H_1\,\triangleleft\,\cdots\,\triangleleft H_m = G.
$
Note that $H_{j+1}/H_j$ is necessarily cyclic in this case for each $j$.
Given an arbitrary collection of generators for a solvable group $G$, a
polynomial-length sequence $g_1,\ldots,g_m$ as above can be found via a
(classical) Monte~Carlo algorithm in polynomial time \cite{BabaiC+95}
(discussed in more detail below).
It is important to note that we allow the possibility that $H_j = H_{j+1}$
for some values of $j$ in reference to this claim.

We will be working in the general context of black-box groups, which we
now discuss.
In a black-box group, each elements is uniquely encoded by some binary string,
and we have at our disposal a black-box (or group oracle) that performs
the group operations on these encodings at unit cost.
For a given black-box group, all of the encodings are of a fixed length $n$,
which is the {\em encoding length}.
Thus, a black-box group with encoding length $n$ has order bounded above by
$2^n$.
Note that not every binary string of length $n$ necessarily corresponds to a
group element, and we may imagine that our group oracle has some arbitrary
behavior given invalid encodings.
(Our algorithms will never query the oracle for invalid group element
encodings given valid input elements).
When we say that a particular group or subgroup is given (to some algorithm),
we mean that a set of strings that generate the group or subgroup is given.
Note that every subgroup of a black-box group with encoding length $n$ has a
length $O(n^2)$ description.

Since we will be working with quantum circuits, we must describe black-box
groups in this setting.
Corresponding to a given black-box group $G$ with encoding length $n$ is a
quantum gate $U_G$ acting on $2n$ qubits as follows:
$
U_G\ket{g}\ket{h} = \ket{g}\ket{gh}.
$
Here we assume $g$ and $h$ are valid group elements---in case any invalid
encoding is given, $U_G$ may act in any arbitrary way so long as is remains
reversible.
The inverse of $U_G$ acts as follows:
$
U_G^{-1}\ket{g}\ket{h} = \ket{g}\ket{g^{-1}h}.
$
When we say that a quantum circuit has access to a group oracle for $G$, we
mean that the circuit may include the gates $U_G$ and $U_G^{-1}$ for some
$U_G$ as just described.
More generally, when we are discussing uniformly generated families of
quantum circuits, a group oracle corresponds to an infinite sequence of
black-box groups $G_1,G_2,\ldots$ (one for each encoding length), and we allow
each circuit in the uniformly generated family to include gates of the
form $U_{G_n}$ and $U_{G_n}^{-1}$ for the appropriate value of~$n$.

As noted by Mosca \cite{Mosca99}, the gates $U_G$ and $U_G^{-1}$ above can be
approximated efficiently if we have a single gate $V_G$ acting as follows on
$3n$ qubits:
$V_G\ket{g}\ket{h}\ket{x} = \ket{g}\ket{h}\ket{x\oplus g h}$,
again where we assume $g$ and $h$ are valid group elements (and $x$ is
arbitrary).
Here, $x\oplus gh$ denotes the bitwise exclusive or of the string $x$ and
the string encoding the group element $gh$.
This claim follows from the fact that given the gate $V_G$, we may find the
order of any element $g$ using Shor's algorithm, from which we may find the
inverse of $g$.
Once we have this, techniques in reversible computation due to
Bennett~\cite{Bennett73} allow for straightforward simulation of $U_G$ and
$U_G^{-1}$.
Since it is simpler to work directly with the gates $U_G$ and $U_G^{-1}$,
however, we will assume that these are the gates made available for a given
black-box group.

Now we return to the topic of solvable groups, and review some known facts
about solvable groups in the context of black-box groups.
First, with respect to any given group oracle, if we are given generators
$g_1,\ldots,g_m$ of encoding length $n$, it is possible to test whether
$G = \langle g_1,\ldots,g_m\rangle$ is solvable via a polynomial time
(in $nm$) Monte Carlo algorithm~\cite{BabaiC+95}.
Moreover, the same algorithm can be used to construct (with high probability)
generators $g_1^{(j)},\ldots,g_k^{(j)}$, for $j=0,\ldots,n$ and where
$k = O(n)$, such that $G^{(j)} = \langle g_1^{(j)},\ldots,g_k^{(j)}\rangle$
(so that testing solvability can be done by verifying that
$g_1^{(n)},\ldots,g_k^{(n)}$ are each the identity element).
At this point we notice (under the assumption that $G$ is solvable) that by
relabeling the elements
\[
g_1^{(n-1)},\ldots,g_k^{(n-1)},g_1^{(n-2)},\ldots,g_k^{(n-2)},\ldots,
g_1^{(0)},\ldots,g_k^{(0)},
\]
as $h_1,\ldots,h_{kn}$ (in the order given) we have the following.
If $H_j = \langle h_1,\ldots,h_j\rangle$ for $j=0,\ldots,kn$, then
$\{1\}=H_0\,\triangleleft H_1\,\triangleleft\,\cdots\,\triangleleft H_{kn}=G$.
This follows from the fact that $G^{(j)}\triangleleft G^{(j-1)}$ for each
$j$, and further that $G^{(j-1)}/G^{(j)}$ is necessarily abelian.
The fact that each factor group $H_j/H_{j-1}$ is cyclic will be important
for our quantum algorithm in the next section.

The problem of computing the order of a group cannot be solved classically in
polynomial time in the black-box setting even for abelian (and therefore for
solvable) groups~\cite{BabaiS84}.


\section{Finding the orders of solvable groups}
\label{sec:algorithm}

In this section we describe our quantum algorithm for finding the order of a
given solvable black-box group $G$ and preparing a uniform superposition over
the elements of $G$.

We assume we have elements $g_1,\ldots,g_m\in G$ such that if we define
$H_j = \langle g_1,\ldots,g_j\rangle$ for each $j$, then
$\{1\} = H_0\,\triangleleft H_1\,\triangleleft\,\cdots\,\triangleleft H_m = G$.
Note that we allow the possibility that $H_j = H_{j+1}$ for some values
of $j$.
The existence of such a chain is equivalent to the solvability of $G$, and
given an arbitrary collection of generators of $G$ such a sequence can
be found via a Monte Carlo algorithm in polynomial time as discussed in the
previous section.
Calculation of the orders of the factor groups in this chain reveals the order
of $G$; if
\begin{equation}
r_1 = |H_1/H_0|,\;\;r_2 = |H_2/H_1|,\;\; \ldots,\; r_m = |H_m/H_{m-1}|,
\label{eq:normal_chain}
\end{equation}
then $|G| = \prod_{j=1}^mr_j$.

The calculation of the orders of the factor groups is based on the following
idea.
Suppose we have several copies of the state $\ket{H}$ for some subgroup $H$ of
$G$, where $\ket{H}$ denotes the state that is a uniform superposition over
the elements of $H$:
\[
\ket{H} = \frac{1}{\sqrt{|H|}}\sum_{h\in H}\ket{h}.
\]
Then using a simple modification of Shor's order finding algorithm we may find
the {\em order of $g$ with respect to $H$}, which is the smallest positive
integer $r$ such that $g^r\in H$, for any $g\in G$.
In case $H = \langle g_1,\ldots,g_{j-1}\rangle$ and $g = g_j$ for some
$j$, this order is precisely $r_j = |H_j/H_{j-1}|$.

Since this requires that we have several copies of $\ket{H_{j-1}}$ in order to
compute each $r_j$, we must demonstrate how the state $\ket{H_{j-1}}$ may be
efficiently constructed.
In fact, the construction of the states $\ket{H_0},\ket{H_1},\ldots$ is done
in conjunction with the computation of $r_1,r_2,\ldots$; in order to
prepare several copies of $\ket{H_j}$ it will be necessary to compute $r_j$,
and in turn these copies of $\ket{H_j}$ are used to compute $r_{j+1}$.
This continues up the chain until $r_m$ has been computed and $\ket{H_m}$
has been prepared.
More specifically, we will begin with a large (polynomial) number of copies
of $\ket{H_0}$ (which are of course trivial to prepare), use some relatively
small number of these states to compute $r_1$, then convert the rest of the
copies of $\ket{H_0}$ to copies of $\ket{H_1}$ using a procedure described
below (which requires knowledge of $r_1$).
We continue up the chain in this fashion, for each $j$ using a relatively
small number of copies of $\ket{H_{j-1}}$ to compute $r_j$, then converting
the remaining copies of $\ket{H_{j-1}}$ to copies of $\ket{H_j}$.

In subsections~\ref{sec:order} and \ref{sec:create} we discuss the two
components (computing the $r_j$ values and converting copies of
$\ket{H_{j-1}}$ to copies of $\ket{H_j}$) individually, and in
subsection~\ref{sec:together} we describe the main algorithm that combines
the two components.
The following notation will be used in these subsections.
Given a finite group $G$ and a subgroup $H$ of $G$, for each element $g\in G$
define $r_H(g)$ to be the smallest positive integer $r$ such that $g^r\in H$
(which we have referred to as the order of $g$ with respect to $H$).
For any positive integer $m$ and $k\in\mathbb{Z}_m$ we write $e_m(k)$
to denote $e^{2\pi i k/m}$.
Finally, for any finite set $S$ we write
$\ket{S} = |S|^{-1/2}\sum_{g\in S}\ket{g}$.


\subsection{Finding orders with respect to a subgroup}
\label{sec:order}

Our method for computing the order of an element $g$ with respect to a
subgroup $H$ (i.e., computing the $r_j$ values) is essentially Shor's
(order finding) algorithm, except that we begin with one of the registers
initialized to $\ket{H}$, and during the algorithm this register is reversibly
multiplied by an appropriate power of $g$.
In short, initializing one of the registers to $\ket{H}$ gives us an easy way
to work over the cosets of $H$, the key properties being (i) that the
states $\ket{g^i H}$ and $\ket{g^j H}$ are orthogonal whenever $g^i$ and $g^j$
are elements in different cosets of $H$ (and of course
$\ket{g^i H} = \ket{g^j H}$ otherwise), and (ii) for Shor's algorithm we will
not need to be able to recognize which coset we are in (or even look at
the corresponding register at all) to be able to compute the order of $g$
with respect to $H$ correctly.

Now we describe the method in more detail.
However, since the analysis is almost identical to the analysis of Shor's
algorithm, we will not discuss the analysis in detail and instead refer the
reader to Shor~\cite{Shor97} and to other sources in which analyses of
closely related techniques are given in detail~\cite{CleveE+98, Kitaev95}.

We assume we are working over a black-box group $G$ with encoding length $n$,
and that a quantum register $\mathbf{R}$ has been initialized to state
$\ket{H}$ for $H$ some subgroup of $G$.
For given $g$ we are trying to find $r = r_H(g)$, which is the smallest
positive integer such that $g^r\in H$.
Let $\mathbf{A}$ be a quantum register whose basis states correspond to
$\mathbb{Z}_N$ for $N$ to be chosen later, and assume $\mathbf{A}$ is
initialized to state $\ket{0}$.

Similar to Shor's algorithm, we (i) perform the quantum Fourier transform
modulo $N$ ($\op{QFT}_N$) on $\mathbf{A}$, (ii) reversibly left-multiply the
contents of $\mathbf{R}$ by $g^a$, for $a$ the number contained in
$\mathbf{A}$, and (iii) perform $\op{QFT}_N^{\dagger}$ on $\mathbf{A}$.
Multiplication by $g^a$ can easily be done reversibly in polynomial
time using the group oracle along with repeated squaring.
The state of the pair $(\mathbf{A},\mathbf{R})$ is now
\[
\frac{1}{N}\sum_{a\in\mathbb{Z}_N}\sum_{b\in\mathbb{Z}_N}
e_N(-ab)\ket{b}\ket{g^a H}.
\]
Observation of $\mathbf{A}$ yields some value $b\in\mathbb{Z}_N$; we will have
with high probability that $b/N$ is a good approximation to $k/r$ (with
respect to ``modulo 1'' distance), where $k$ is randomly distributed in
$\mathbb{Z}_r$.
Assuming $N$ is sufficiently large, we may find relatively prime integers
$u$ and $v$ such that $u/v = k/r$ with high probability via the continued
fraction method---choosing $N = 2^{2n+O(\log(1/\varepsilon))}$ allows us to
determine $u$ and $v$ with probability $1 - \varepsilon$.
Now, to find $r$, we repeat this process $O(\log (1/\varepsilon))$ times
and compute the least common multiple of the $v$ values, which yields $r$ with
probability at least $1-\varepsilon$.


\subsection{Creating uniform superpositions over subgroups}
\label{sec:create}

Next we describe how several copies of the state $\ket{H}$ may be converted
to several copies of the state $\ket{\langle g\rangle H}$.
It is assumed that $g$ normalizes $H$ (i.e., $gH = Hg$, implying that
$\langle g\rangle H$ is a group and that $H \triangleleft \langle g\rangle H$)
and further that $r = r_H(g) = |\langle g\rangle H/H|$ is known.
For the main algorithm this corresponds to converting the copies of
$\ket{H_{j-1}}$ to copies of $\ket{H_j}$.
We note that this is the portion of the algorithm that apparently
requires the normal subgroup relations in (\ref{eq:normal_chain}), as the
assumption that $g$ normalizes $H$ is essential for the method.

Specifically, for sufficiently large $l$, $l$ copies of $\ket{H}$ are
converted to $l-1$ copies of $\ket{\langle g\rangle H}$ with high probability;
the procedure fails to convert just one of the copies.
We assume that we have registers $\mathbf{R}_1,\ldots,\mathbf{R}_l$, each in
state $\ket{H}$.
Let $\mathbf{A}_1,\ldots,\mathbf{A}_l$ be registers whose basis states
correspond to $\mathbb{Z}_r$, and assume $\mathbf{A}_1,\ldots,\mathbf{A}_l$
are each initialized to $\ket{0}$.
For each $i = 1,\ldots,l$ do the following:
(i) perform $\op{QFT}_r$ on register $\mathbf{A}_i$,
(ii)~(reversibly) left-multiply the contents of $\mathbf{R}_i$ by $g^{a_i}$,
where $a_i$ denotes the contents of $\mathbf{A}_i$, and
(iii)~again perform $\op{QFT}_r$ on $\mathbf{A}_i$.
Each pair $(\mathbf{A}_i,\mathbf{R}_i)$ is now in the state
\[
\frac{1}{r}\sum_{a_i\in\mathbb{Z}_r}\sum_{b_i\in\mathbb{Z}_r}
e_r(a_i b_i)\ket{b_i}\ket{g^{a_i}H}.
\]
Now, measure $\mathbf{A}_1,\ldots,\mathbf{A}_l$, denoting the results by
$b_1,\ldots,b_l$.
Let $\ket{\psi_i}$ denote the resulting (normalized) state of $\mathbf{R}_i$
for each $i$, i.e.,
\[
\ket{\psi_i} = \frac{1}{\sqrt{r}}\sum_{a_i\in\mathbb{Z}_r}e_r(a_i b_i)
\ket{g^{a_i}H}.
\]

Now we hope that at least one of the values $b_i$ is relatively prime to $r$;
this fails to happen with probability at most $\varepsilon$ whenever
$l\in\Omega((\log\log r)(\log 1/\varepsilon))$.
Assuming we are in this case, choose $k$ such that $b_k$ is relatively prime
to $r$.
We will use $\ket{\psi_k}$ to ``correct'' the state in each of the remaining
registers $\mathbf{R}_i$, $i\not=k$, by doing
the following: reversibly multiply the contents of $\mathbf{R}_k$ by $f^c$,
where $f$ denotes the group element contained in $\mathbf{R}_i$ and $c$ is any
integer satisfying $c\equiv b_i b_k^{-1}\,(\bmod\,r)$.
We claim at this point that $\mathbf{R}_i$ contains the state
$\ket{\langle g\rangle H}$ and $\mathbf{R}_k$ is unchanged (i.e., still
contains $\ket{\psi_k}$).
To see this, consider an operator $M_{g^jh}$ that multiplies the contents of
$\mathbf{R}_k$ by $g^jh$ (for arbitrary $h\in H$).
As $g$ normalizes $H$ we have
\[
M_{g^jh}\ket{\psi_k}
\:=\:\frac{1}{\sqrt{r}}\sum_{a_k\in\mathbb{Z}_r}e_r(a_k b_k)\ket{g^{j+a_k}H}
\:=\:\frac{1}{\sqrt{r}}\sum_{a_k\in\mathbb{Z}_r}e_r((a_k-j) b_k)\ket{g^{a_k}H}
\:=\:e_r(-jb_k)\ket{\psi_k},
\]
which shows that the state $\ket{\psi_k}$ is an eigenvector of $M_{g^jh}$ with
associated eigenvalue $e_r(-jb_k)$.
Thus, after performing the above multiplication, the state of the pair
$(\mathbf{R}_i,\mathbf{R}_k)$ is
\begin{eqnarray*}
\frac{1}{\sqrt{r|H|}}
\sum_{a_i\in\mathbb{Z}_r}
\sum_{h\in H}
e_r(a_i b_i)
\ket{g^{a_i}h}
M_{(g^{\smash{a\hspace{-.8pt}{}_i}}h)^c}\ket{\psi_k}
& = &
\frac{1}{\sqrt{r|H|}}
\sum_{a_i\in\mathbb{Z}_r}
\sum_{h\in H}
e_r(a_i b_i -  a_i b_i b_k^{-1}b_k)
\ket{g^{a_i}h}
\ket{\psi_k}\\
& = &
\frac{1}{\sqrt{r|H|}}
\sum_{a_i\in\mathbb{Z}_r}
\sum_{h\in H}
\ket{g^{a_i}h}
\ket{\psi_k}\\
& = &
\ket{\langle g\rangle H}\,\ket{\psi_k}.
\end{eqnarray*}
This procedure is repeated for each value of $i\not=k$ and then
$\mathbf{R}_k$ is discarded; this results in $l-1$ copies of
$\ket{\langle g\rangle H}$ as desired.

It should be noted that it is not really necessary that one of the $b_i$
values is relatively prime to $r$, but a more complicated procedure is
necessary in the more general case.
Since we already have a polynomial-time algorithm without the more complicated
procedure, we will not discuss it further.


\subsection{The main algorithm}
\label{sec:together}

As above, we assume we have elements $g_1,\ldots,g_m\in G$ such that for
$H_j = \langle g_1,\ldots,g_j\rangle$ for each $j$, we have
$\{1\}=H_0\,\triangleleft\,H_1\,\triangleleft\,\cdots\,\triangleleft\,H_m = G$.
Defining $r_j = r_{H_{j-1}}(g_j) = |H_j/H_{j-1}|$ for each $j$ we have
$|G| = \prod_{j=1}^m r_j$.
Consider the algorithm in Figure~\ref{fig:algorithm}.
Here, $k$ is a parameter to be chosen later.

\begin{figure}[ht]
\noindent\hrulefill\vspace{-3mm}

\begin{tabbing}
8\=888\=8888\=8888\=\+\kill

Prepare $k(m+1)$ copies of the state $\ket{H_0}$, where $H_0=\{1\}$.\\[2mm]

Do the following for $j = 1,\ldots, m$:\\[2mm]

\> Using $k-1$ of the copies of $\ket{H_{j-1}}$, compute $r_j=r_{H_{j-1}}(g_j)$
(and discard these $k-1$ states).\\[2mm]

\> Use one of the copies of $\ket{H_{j-1}}$ to convert the remaining copies of
$\ket{H_{j-1}}$ to copies of $\ket{H_j}$.\\[2mm]

End of for loop.\\[2mm]

Output $\prod_{j=1}^mr_j$.\\[-9mm]
\end{tabbing}
\noindent\hrulefill
\caption{Algorithm to compute the order of a solvable group $G$}
\label{fig:algorithm}
\end{figure}

It is clear that the algorithm operates correctly assuming that each
evaluation of $r_j$ is done without error, and that the copies of
$\ket{H_{j-1}}$ are converted to copies of $\ket{H_j}$ without error on
each iteration of the loop.
To have that the algorithm works correctly with high probability in general,
we must simply choose parameters so that the error in all of these steps is
small.
If we want the entire process to work with probability of
error less than $\varepsilon$, we may perform the computations of each of the
$r_j$ values such that each computation errs with probability at most
$\varepsilon/(2m)$, and for each $j$ the copies of $\ket{H_{j-1}}$ are
converted to copies of $\ket{H_j}$ with error at most $\varepsilon/(2m)$.
Thus, choosing $k = O((\log n)(\log m/\varepsilon))$ suffices.
In polynomial time we may therefore achieve an exponentially small probability
of error by choosing $k$ polynomial in $n$ and computing the $r_j$ values with
sufficient accuracy.


\section{Other problems}
\label{sec:other}

In this section we discuss other problems regarding solvable groups that
can be solved in quantum polynomial time with the help of our main algorithm.
First we discuss membership testing and other problems that easily reduce to
computing order.
We then we discuss the general technique for computing over factor groups of
solvable groups.


\subsection{Membership testing and simple reductions to order finding}

Suppose we are given elements $g_1,\ldots,g_k$ and $h$ in some black-box group
with encoding length $n$.
Clearly $h\in\langle g_1,\ldots,g_k\rangle$ if and only if
$|\langle g_1,\ldots,g_k\rangle| = |\langle g_1,\ldots,g_k,h\rangle|$.
Thus, if $\langle g_1,\ldots,g_k,h\rangle$ is solvable, then the question of
whether $h\in\langle g_1,\ldots,g_k\rangle$ can be computed in quantum
polynomial time.
Since there is a classical algorithm for testing solvability, it is really
only necessary that $\langle g_1,\ldots,g_k\rangle$ is solvable; if
$\langle g_1,\ldots,g_k\rangle$ is solvable but
$\langle g_1,\ldots,g_k,h\rangle$ is not, then clearly
$h\not\in\langle g_1,\ldots,g_k\rangle$.

Several other problems reduce to order computation or membership testing in
solvable groups.
A few examples are testing whether a given solvable group is a subgroup of
another (given $g_1,\ldots,g_k$ and $h_1,\ldots,h_l$, is it the case that
$\langle h_1,\ldots,h_l\rangle \leq \langle g_1,\ldots,g_k\rangle$?), testing
equality of two solvable groups (given $g_1,\ldots,g_k$ and $h_1,\ldots,h_l$,
is it the case that
$\langle g_1,\ldots,g_k\rangle=\langle h_1,\ldots,h_l\rangle$?), and testing
whether a given group is a normal subgroup of a given solvable group (given
$g_1,\ldots,g_k$ and $h_1,\ldots,h_l$, do we have
\mbox{$\langle h_1,\ldots,h_l\rangle \triangleleft\langle g_1,\ldots,
g_k\rangle$}?).
To determine whether $\langle h_1,\ldots,h_l\rangle$ is a subgroup of
$\langle g_1,\ldots,g_k\rangle$, we may simply test that 
$|\langle h_1,\ldots,h_l,g_1,\ldots,g_k\rangle|
= |\langle g_1,\ldots,g_k\rangle|$
(or we may test that each $h_j$ is an element of
$\langle g_1,\ldots,g_k\rangle$ separately), to test equality we verify that
$\langle g_1,\ldots,g_k\rangle \leq \langle h_1,\ldots,h_l\rangle$
and $\langle h_1,\ldots,h_l\rangle \leq \langle g_1,\ldots,g_k\rangle$,
and to test normality we may verify that
$g_i^{-1} h_j g_i \in\langle h_1,\ldots,h_l\rangle$ for each $i$ and $j$ (as
well as $\langle h_1,\ldots,h_l\rangle \leq \langle g_1,\ldots,g_k\rangle$).
See Babai~\cite{Babai92} for more examples of problems reducing to order
computation.

In another paper \cite{Watrous00} we have shown that there exist succinct
{\em quantum} certificates for various group-theoretic properties, including
the property that a given integer divides the order of a group (i.e.,
given an integer $d$ and generators $g_1,\ldots,g_k$ in some black-box group,
where $G = \langle g_1,\ldots,g_k\rangle$ is not necessarily solvable,
verify that $d$ divides $|G|$).
We note here that our quantum algorithm for calculating orders of solvable
groups can be used to prove the existence of succinct {\em classical}
certificates for this property.
Suppose we are given $d$ and $g_1,\ldots,g_k$ as above.
Then a classical certificate for the property that $d$ divides $|G|$ may
consist of descriptions of $p$-subgroups of $G$ for the primes $p$ dividing
$d$.
More specifically, suppose $d = p_1^{a_1}\cdots p_m^{a_m}$ for distinct
primes $p_1,\ldots,p_m$.
Then for each prime power $p_j^{a_j}$, the certificate will include a
description of some subgroup of $G$ having order $p_j^{a_j}$.
If $p_j^{a_j}$ indeed divides $|G|$ there will exist such a subgroup, which
is necessarily solvable since all groups of prime power order are solvable.
Thus, the order of each given $p$-subgroup can be found using the order
calculation algorithm.
Since $G$ is not necessarily solvable, however, testing that each of the given
$p$-subgroups is really a subgroup of $G$ might not be possible with our
algorithm.
However, the certificate may also include proofs of membership for each of the
generators of the $p$-subgroups in $G$.
(See Babai and Szemer\'edi~\cite{BabaiS84} for details on proofs of
membership.)


\subsection{Computing over abelian factor groups}

In the case of abelian black-box groups, many group-theoretic problems can be
solved in polynomial time on a quantum computer.
For instance, given generators $g_1,\ldots,g_k$ for an abelian black-box group
$G$ with encoding length $n$, in quantum polynomial time we may compute prime
powers $q_1,\ldots,q_m$ such that
$G \,\cong\, \mathbb{Z}_{q_1}\times\cdots\times\mathbb{Z}_{q_m}$.
Furthermore, there exists an isomorphism\linebreak
$\theta:G\rightarrow\mathbb{Z}_{q_1}\times\cdots\times\mathbb{Z}_{q_m}$ such
that for any $h\in G$, $\theta(h)$ may be computed in time polynomial in $n$.
Consequently, computing the order of an abelian group, testing isomorphism
of abelian groups, and several other problems can be performed in quantum
polynomial time \cite{CheungM00,Hoyer00,Mosca99}.

We may apply these techniques for problems about abelian groups to problems
about solvable groups by working over factor groups.
To illustrate how this may be done, consider the following problem.
Suppose we have a solvable group $G$ given by generators $g_1,\ldots,g_k$,
and furthermore that we have generators $h_1,\ldots,h_l$ for a normal
subgroup $H$ of $G$ such that $G/H$ is abelian.
We may hope to determine the structure of $G/H$ using the technique for
abelian groups mentioned above, i.e., we wish to compute prime powers
$q_1,\ldots,q_m$ such that
$G/H\,\cong\,\mathbb{Z}_{q_1}\times\cdots\times\mathbb{Z}_{q_m}$.
However, a complication arises since we do not have unique classical
representations for elements of $G/H$, and so we cannot apply the technique
directly.
Instead, we will rely on the fact that we may efficiently construct copies
of the state $\ket{H}$ in polynomial time in order to work over the factor
group $G/H$.
Assume that $r_1 = \op{order}(g_1),\,\ldots,\,r_k = \op{order}(g_k)$ have
already been computed using Shor's algorithm, and let
$N = \op{lcm}(r_1,\ldots,r_k)$.
The algorithm described in Figure~\ref{fig:factor_group} will allow us to
solve the problem.
\begin{figure}[ht]
\noindent\hrulefill\vspace{-2mm}

\begin{tabbing}
8\=888\=8888\=8888\=\+\kill

Prepare register $R$ in state $\ket{H}$ using the algorithm from
Section~\ref{sec:algorithm}.\\[2mm]

Initialize registers $A_1,\ldots,A_k$ each in state
$\frac{1}{\sqrt{N}}\sum_{a = 0}^{N-1}\ket{a}$.\\[2mm]

Reversibly (left-)multiply the contents of register $R$ by
$g_1^{a_1}\cdots g_k^{a_k}$, where each $a_j$ denotes the\\
contents of register $A_j$.\\[2mm]

For $j = 1,\ldots,k$, perform the quantum Fourier transform modulo $N$ on
register $A_j$.\\[2mm]

Observe $A_1,\ldots,A_k$ (in the computational basis).\\[-9mm]

\end{tabbing}
\noindent\hrulefill
\caption{Quantum subroutine used for determining the structure of $G/H$.}
\label{fig:factor_group}
\end{figure}

To analyze the algorithm, define a mapping
$f:\mathbb{Z}_N^k\rightarrow G/H$ as
$f(a_1,\ldots,a_k) = g_1^{a_1}\cdots g_k^{a_k}H$.
The mapping $f$ is a homomorphism with
$\op{ker}(f)=\{(a_1,\ldots,a_k)\in\mathbb{Z}_N^k\,
|\,g_1^{a_1}\cdots g_k^{a_k}\in H\}$.
Define
\[
\op{ker}(f)^{\perp} = \left\{(b_1,\ldots,b_k)\in\mathbb{Z}_N^k\,
\left|\,\sum_{j=1}^k a_j b_j \equiv 0 \:(\op{mod}\:N)\:
\mbox{for all $(a_1,\ldots,a_k)\in\op{ker}(f)$}\right.\right\}.
\]
We have that $\op{ker}(f)^{\perp} \cong G/H$, and in fact
$f$ is an isomorphism when restricted to $\op{ker}(f)^{\perp}$.
A straightforward analysis reveals that observation of $A_1,\ldots,A_k$ will
give a random element in $\op{ker}(f)^{\perp}$.

Thus, running the algorithm in Figure~\ref{fig:factor_group} $O(k)$ times
results in a generating set for $\op{ker}(f)^{\perp}$ with high probability.
Letting $B$ be a matrix whose columns are the randomly generated elements
of $\op{ker}(f)^{\perp}$, we may determine the numbers $q_1,\ldots,q_m$
in polynomial time by computing the Smith normal form of $B$
(see Kannan and Bachem \cite{KannanB79} and Hafner and
McCurley \cite{HafnerM91} for polynomial-time algorithms for computing Smith
normal forms).


This method for working over factor groups can be applied to other problems.
In general, we may represent elements in a factor group $G/H$ by quantum
states of the form $\ket{gH}$.
Two states $\ket{gH}$ and $\ket{g'H}$ are of course identical whenever
$gH=g'H$, and are orthogonal otherwise.
Multiplication and inversion of such states works as expected---for
$U_G$ as in Section~\ref{sec:preliminaries} we have
$U_G\ket{gH}\ket{g'H} = \ket{gH}\ket{gg'H}$ and
$U_G^{-1}\ket{gH}\ket{g'H} = \ket{gH}\ket{g^{-1}g'H}$.
(This requires $H\triangleleft G$.)
Hence this gives us a natural way to represent elements of factor groups by
quantum states.


\section{Conclusion}
\label{sec:conclusion}

We have given a polynomial-time quantum algorithm for calculating the order
and preparing a uniform superposition over a given solvable group, and shown
how this algorithm may be used to solve other group-theoretic problems
regarding solvable groups in polynomial time.

There are several other problems for solvable black-box groups that we
do not have polynomial-time algorithms for.
Examples include Group Intersection (given generating sets for two
subgroups of a solvable black-box group, do the subgroups have a nontrivial
intersection?) and Coset Intersection (defined similarly).
See Arvind and Vinodchandran \cite{ArvindV97} and Babai \cite{Babai92} for
more examples of group-theoretic problems we may hope to solve in quantum
polynomial time in the solvable black-box group setting.

Another interesting question is whether there exist polynomial-time quantum
algorithms for similar problems for arbitrary (not necessarily solvable) finite
groups.
Can our methods be extended to non-solvable groups, and if so, to what extent?
One possible approach to the particular problem of calculating group order is
to try and develop an algorithm to find generators for the Sylow subgroups of
the given group, and to run our algorithm on these subgroups
(which are necessarily solvable).


\subsection*{Acknowledgments}

I thank Eric Bach, Richard Cleve, Alexei Kitaev, Michele Mosca, and
Miklos Santha for helpful comments and suggestions.


\bibliographystyle{plain}


\end{document}